\documentclass[lettersize,onecolumn]{IEEEtran}
\usepackage{amsmath,amsfonts,amsthm}
\usepackage{algorithm}
\usepackage{array}
\usepackage{algpseudocode}
\usepackage[caption=false,font=normalsize,labelfont=sf,textfont=sf]{subfig}
\usepackage{textcomp}
\usepackage{stfloats}
\usepackage{url}
\usepackage{verbatim}
\usepackage{graphicx}
\usepackage{cite}
\usepackage[colorlinks=true,linkcolor=blue]{hyperref}

\usepackage[normalem]{ulem}

\newtheorem{theorem}{Theorem}[section]
\newtheorem{lemma}{Lemma}[section]

\newtheorem{assumption}{Assumption}[section]
\newtheorem{example}{Example}[section]

\newcommand{\Sig}{\mathbf{\Sigma}}

\newcommand{\Eb}{\mathbf{E}}

\newcommand{\E}{\mathbf{E}}

\newcommand{\Lb}{\mathbf{L}}

\newcommand{\RR}{\mathcal{R}}

\newcommand{\F}{\mathcal{F}}
\newcommand{\sig}{\sigma}

\newcommand{\gam}{\gamma}

\newcommand{\al}{\alpha}

\newcommand{\ep}{\epsilon}

\newcommand{\Xc}{\mathcal{X}}

\newcommand{\Zb}{\mathbf{Z}}

\newcommand{\de}{\delta}

\newcommand{\xc}{\mathbf{x}}

\newcommand{\Ab}{\mathbf{A}}

\newcommand{\Qb}{\mathbf{Q}}

\let\tilde\widetilde

\usepackage{todonotes, ulem}

\usepackage{booktabs}

\begin{document}

\title{Sequential Change Point Detection with FDR Control in Reconfigurable Sensor Networks}

\author{Seungwon Lee,  Yunxiao Chen, Xiaoou Li
\thanks{Seungwon Lee is with the School of Statistics, University of Minnesota, Minneapolis, MN, USA (e-mail:lee03949@umn.edu)
\\ Yunxiao Chen is with Department of Statistics, London School of Economics and Political Science, London, U.K. (e-mail:Y.Chen186@lse.ac.uk)
\\ Xiaoou Li is with School of Statistics, University of Minnesota, Minneapolis, MN, USA (e-mail:lixx1766@umn.edu)}}

\maketitle

\begin{abstract}
This paper investigates sequential change-point detection in reconfigurable sensor networks. In this problem, data from multiple sensors are observed sequentially. Each sensor can have a unique change point, and the data distribution differs before and after the change. We aim to detect these changes as quickly as possible once they have occurred while controlling the false discovery rate at all times. Our setting is more realistic than traditional settings in that (1) the set of active sensors -- i.e., those from which data can be collected -- can change over time through the deactivation of existing sensors and the addition of new sensors and (2) dependencies can occur 
both between sensors and across time points. We propose powerful e-value-based detection procedures that control the false discovery rate uniformly over time. Numerical experiments demonstrate that, with the same false discovery rate target, our procedures achieve superior performance compared to existing methods, exhibiting lower false non-discovery rates and reduced detection delays.
\end{abstract}


\section{Introduction} \label{Introduction}
\IEEEPARstart{T}{he} classical 
sequential change detection problem
concerns data from a single data stream. The aim is to detect the change point as quickly as possible, 
subject to the constraint that false detection occurring before the true change point is rare.
Fundamental early developments include  
\cite{page1954continuous, roberts1966comparison, shiryaev1963optimum, lorden1971procedures}.  
The optimality of these procedures has been extensively studied under various statistical criteria and problem settings; see, e.g., \cite{baron2006asymptotic, feinberg2006quickest, shiryaev2007optimal, nikiforov1997two, moustakides2014multiple, lai1998information,  pollak1985optimal}. We refer the readers to \cite{lai2010sequential} and \cite{tartakovsky2014} for a review of the methods and theory for the classical sequential change detection problem.

In recent years, with the emergence of large-scale and complex data, change detection based on multi-stream data has received wide attention. 
The settings for multi-stream change detection may be classified into two categories. The first category addresses settings with one or multiple change points affecting all or some data streams, where the objective is to detect these common change points by leveraging information across the streams. 
Various methods have been developed under settings of this category, including  \cite{tartakovsky2008asymptotically,zarrin2009cooperative,mei2011quickest,egea2017space,chan2017optimal,mei2010efficient,xie2013sequential,fellouris2016second,chen2015graph,chen2019sequential,song2024practical}. The second category, which will be the focus of the current work,  concerns
settings in which the change points are 
stream-specific. Under such settings, multiple streams may have changed at each time, and the goal is to detect the changed streams while controlling for a certain compound risk that measures the overall quality of change detection across streams, such as a False Discovery Rate (FDR). These settings have received increasing attention in recent years. Various methods have been developed under specific settings, and applications to spectral sensing for cognitive radios and item pool quality control in standardized testing have been explored; see, e.g., \cite{chen2016,chen2020false, chen2022, lu2022, chen2023compound,dandapanthula2025}.  

Unfortunately, the settings explored in existing works on stream-specific changes are overly restricted and do not fully reflect the complexity of real-world situations. One restriction involves data dependence, as most existing works require the data to be independent across streams and time points. Another restriction comes from the partial observation of the streams. In some applications, the data streams detected to have changed need to be deactivated, after which data can no longer be collected from these streams. In addition, new data streams may be added to the monitoring system over time. Data from these streams can only be observed once the streams are included in the monitoring system. 
The adaptivity of the active streams (i.e., streams from which we can observe data) introduces additional dependence on the observed data, making the control of the compound risk more challenging. 
Specifically, \cite{chen2016,chen2020false, chen2022, lu2022, chen2023compound} concern settings where detected streams are deactivated while data are independent across streams and time. These works, except for \cite{chen2022}, do not allow new streams to be added during the monitoring process. \cite{dandapanthula2025} propose an e-value-based approach for FDR control that works for dependent data. However, they require the data to be fully observed, i.e., stream deactivation and addition are not allowed. 

This paper proposes a statistical methodology for detecting stream-specific changes with FDR control. We consider a very general setting, allowing data to be dependent across streams and time, detected streams to be deactivated, and new streams to be added. For ease of exposition, we focus on the context of a reconfigurable sensor network while pointing out that our development can be applied to more general contexts beyond sensor networks, such as the item pool quality control problem in standardized testing. %
Data are collected sequentially for all the sensors in the network, and the network may evolve over time due to the deactivation of detected sensors or the addition of new sensors. To control the false discovery rate, we adopt an e-value-based method, extending the e-BH procedure proposed in \cite{wang2022} to the sequential setting. Although the two methods are closely related, we developed this method independently of the work presented in \cite{dandapanthula2025}, which was posted on arXiv during the preparation of this manuscript. In addition, the current paper considers a more general setting where the active data streams can evolve adaptively over time, while \cite{dandapanthula2025} require the active streams to remain unchanged, which is a special case of ours. 
The key contributions of this paper are as follows:
\begin{enumerate}
\item{\textit{General setting:} The proposed method accommodates complex dependencies across sensors (i.e., streams) and time, substantially relaxing the traditional independence assumptions.}
\item{\textit{Flexible active set control:} We allow dynamic adjustments in the  set of active sensors at each time point under mild constraints. Users can deactivate the detected sensors or keep them in the network. They are also allowed to add 
new sensors to the network, for example, to keep the size of the network unchanged after the deactivation of detected sensors. }

\item{\textit{FDR control:} We introduce three change detection procedures that are theoretically guaranteed to control the FDR under suitable conditions. These procedures include a baseline method based on Shiryaev-Roberts statistic and two boosting methods that improve the detection power of the baseline method. Between the two boosting methods, one operates under the same conditions as the baseline method, while the other requires an additional assumption about the data dependence structure to achieve further improvements.}
\item{\textit{Superior empirical performance:} Our numerical experiments demonstrate that,  
with the same false discovery rate target, our procedures achieve
superior performance compared to existing methods, showing
lower false non-discovery rates and reduced detection delays.}
\end{enumerate}
The remainder of the paper is organized as follows. In Section \ref{problemsetup}, we introduce the setting and performance measures for our multi-stream change detection problem. In Section \ref{methods}, the proposed procedures
are presented, including the baseline procedure and two boosting algorithms. 
 In Section \ref{theoreticalresults}, we establish the theoretical properties of the proposed method and prove uniform control of the FDR across all time points. %
{In Section \ref{discuss2}, we present concrete examples of various change point models and sensor network structures for which the proposed methods are theoretically justified.}
 In Section \ref{Simulationstudies}, we run numerical experiments to evaluate the performance of the proposed methods and compare them with the existing ones.  Finally, we summarize the key findings and discuss the future directions in Section \ref{conclusion}. Proof of our theoretical results is presented in the Appendix.

\section{Problem Setup} \label{problemsetup}
\subsection{Reconfigurable Sensor Networks and Change Detection}
\noindent Let $\Zb_{+}$ be the set of positive integers. Let $k \in \Zb_{+}$ denote the index of a sensor, $t \in \Zb_{+}$ denote the index of time, and $X_{k,t}$ represents the data point collected from the $k$-th sensor at time $t$. We consider a reconfigurable sensor network, where sensors can be activated or deactivated dynamically. Denote $A_t \subset \Zb_{+}$ as the set of indices corresponding to active sensor indices at time $t$. The indices of detected sensors at time $t$ are represented by $D_t$, with the constraint $D_t \subset A_{t-1}$ for all $t \geq 2$. This ensures the detection only applies to previously active sensors. At each time $t$, a data point $X_{k,t}$ is observed if and only if $k \in A_t$. Specifically, define
\begin{equation}
     \Xc_{t} = (X_{k,s})_{k \in A_s, 1 \leq s \leq t}, \text{ and } \Ab_t = (A_s)_{1 \leq s \leq t} 
\end{equation}
The information filtration at time $t$ is then defined as:
\begin{equation}
\F_t = \sig( \Xc_t, \Ab_t ) \text{ for }t\geq 1 \text{ and } \F_0 = \sig( \emptyset, \Omega),
\end{equation}
where $\F_t$ is a $\sigma$-field that includes all observed data, as well as history of the active sensors up to time $t$, and $\F_0$ is the trivial $\sigma$-field.

At each time $t$, a decision maker determines the set of detected sensors $D_{t+1}$ based on the historical information available in $\F_{t}$ (i.e., $D_{t+1}$ is measurable with respect to $\F_{t}$), and an external controller selects the set of active sensors $A_t$. For each sensor $k$, we assume that there is a unique change point $\tau_k$ such that the marginal distribution of $X_{k,t}$ differs before and after $\tau_k$. Our goal is to design a detection rule so that $D_t$ approximates the set of post-change sensors (i.e., $\{k \in A_t : t > \tau_k\}$) so that certain accuracy metric is controlled. Note that we control only $D_t$ rather than $A_t$, since $A_t$ is determined externally. Although we allow dependent observations $X_{k,t}$ and a broad class of processes $A_t$, certain model assumptions are still required. In the next section, we provide detailed model assumptions and introduce performance metrics to quantify the accuracy of the detection rule.

\subsection{Change Point Models}
\noindent For the $k$-th sensor, we assume that there is a deterministic, but unknown change point $\tau_k$ which takes values in $\{0\} \cup \{\infty\} \cup \Zb_{+}$. The change point $\tau_k$ represents the time at which the distribution of the $k$-th sensor transitions from a pre-change regime to a post-change regime. To formalize this, we introduce the following assumption.
\begin{assumption} \label{ass2}
For each $t$, given $A_t$, and $k \in A_{t}$, the conditional distribution of $X_{k,t}$ given the past observations, and active sensor sets has the probability density
\begin{equation}
(X_{k,t} \mid \Xc_{t-1}=\xc_{t-1}, \Ab_{t} = \Ab^{t}) 
\sim
\begin{cases}
f_{0,k,t}(x_{k,t} \mid \xc_{t-1}, \Ab^{t}) & \text{if $t \leq \tau_k$}
\\ f_{1,k,t}(x_{k,t} \mid \xc_{t-1}, \Ab^{t}) & \text{if $t > \tau_k$},
\end{cases}
\end{equation}
for some density functions $f_{0,k,t}$ and $f_{1,k,t}$ with respect to a baseline measure. Furthermore, the support of $f_{1,k,t}(\cdot)$ is either equal or contained within the support of $f_{0,k,t}(\cdot)$ (i.e., $\{x : f_{1,k,t}(x \mid \xc_{t-1}, \Ab^{t}) > 0 \} \subseteq \{x : f_{0,k,t}(x \mid \xc_{t-1}, \Ab^{t}) > 0 \}$).
\end{assumption}
The assumption on the support ensures the likelihood ratio between $f_{1,k,t}(\cdot)$ and $f_{0,k,t}(\cdot)$ is well-defined. We elaborate on the above assumption. First, the density functions $f_{1,k,t}$ and $f_{0,k,t}$  depend on $k$ and $t$, and do not rely on the change points $(\tau_k)_{k \in \Zb_{+}}$. Consequently, the conditional distribution of $X_{k,t}$ given historical information is influenced only by its pre- or post-change status, rather than the specific values of the change points. Second, our analysis requires only the marginal distribution of $X_{k,t}$ conditioned on historical data, without requiring the joint distribution of $X_{k,t}$ across different sensors. Third, we note that this assumption is crucial to ensure the validity of our proposed method. In Section \ref{discuss2}, we give several change point models satisfying this assumption.

We also make the following assumptions on the externally controlled process $\{A_t\}_{t\geq 1}$.
\begin{assumption} \hfill{}
\begin{enumerate} \label{con:restriction}
\item Boundedness: There is a finite constant $K$ such that $\underset{t \geq 1}{\sup}|A_t|\leq K$ almost surely.
\item{Independent sensor addition: The process of selecting active sensors can be random, as long as it does not rely on future observations. That is, for all $k, t \in \Zb_{+}$, $X_{k,t}$ is independent of $A_t$ conditional on $\F_{t-1}$.}
\end{enumerate}
\end{assumption}
We provide some comments on the assumptions regarding the active set $A_t$. Both requirements are mild and easily satisfied in practice. For example, in spectrum sensing, the almost sure upper bound $K$ represents the practical limit on the number of spectra that Secondary Users (SUs)
can monitor at any given time. This constraint arises due to hardware limitations, computational costs, or regulatory restrictions. 
We also note that the process $A_t$ can be dependent on the detection process $(D_s)_{1 \leq s \leq t}$. For example, if $A_t = A_{t-1} \setminus D_t$, this setting reduces to the problem setting introduced in \cite{lu2022, chen2023compound}. Additionally, $A_t$ can be deterministic. In particular, if $A_t = \{1,\cdots,K\}$ for all $t$, all sensors remain active regardless of detection outcomes. {This reduces to the settings in \cite{dandapanthula2025}.}

\subsection{False Discovery Rate and False Non-Discovery Rate}
\noindent Following \cite{chen2023compound,lu2022,chen2022}, we define the false discovery rate (FDR) at time $t$ by
\begin{equation} \label{e:fdr}
\text{FDR}_t = \E \left[ \frac{\sum_{k \in A_t} 1_{(\tau_k \geq t, k \in D_{t+1})}}{\sum_{k\in A_t}1_{(k \in D_{t+1})} \vee 1} \right].
\end{equation}
$\text{FDR}_t$ measures the expected proportion of false detections among all the detections at time $t$. Our primary goal is to design a detection rule to ensure that $\text{FDR}_t\leq \alpha$ for all $t$, for a pre-specified level $\alpha\in (0,1)$. This is equivalent to controlling $\text{maxFDR}=\underset{t\in \Zb_{+}}{\sup} \text{FDR}_t$ to be no greater than $\alpha$. 

If we treat the post-change regime as positive and the pre-change regime as negative, then $\text{FDR}_t$ and $\text{maxFDR}$ measure the false positives of a detection method -- that is, cases where the method incorrectly claims a change during the pre-change regime. To compare methods that control the FDR at the same level, we also need a metric for false negatives. For this purpose, we introduce the aggregated false non-discovery rate (\text{AFNR}) as:
\begin{equation} \label{e:afnr}
\text{AFNR} = \E \left[ \frac{\sum_{t=1}^{\bar{T}-1} \sum_{k \in A_t} 1_{(\tau_k < t, k \notin D_{t+1})} }{ (\sum_{t=1}^{\bar{T}-1} \sum_{k \in A_t} 1_{(k \notin D_{t+1})}) \vee 1} \right].
\end{equation}

Here, $\bar{T}$ is a positive integer serving as the `deadline', which is a user-defined fixed constant. For a comprehensive discussion of the performance metrics for change detection in multiple data streams, we refer the readers to \cite{lu2022}. We develop a change detection method such that $\text{maxFDR}$ is controlled at a target level. Moreover, we develop boosting methods to improve $\text{AFNR}$ of the proposed method.
\section{Methods} \label{methods}

\subsection{Change-Point Detection Procedure} \label{changedetection}
\noindent 
In this section, we first propose a change detection method that controls the maximum false discovery rate (maxFDR) at a pre-specified level $\alpha$, and then explain its rationale. For each time point $i$, 
define the likelihood ratio
\begin{equation}\label{eq:L}
    L_{k,i} = 1_{(k \in A_i)} \frac{f_{1,k,i}(X_{k,i} \mid \Xc_{i-1}, \Ab_{i})}{f_{0,k,i}(X_{k,i} \mid \Xc_{i-1}, \Ab_{i})},
\end{equation}
where $1_{(k \in A_i)}$ is the indicator function that equals $1$ if sensor $k$ is active at time $i$ and $0$ otherwise. %

Define the set of active time indices for sensor $k$ up to time $t$ as $C_{k,t} = \{1 \leq s \leq t \mid k \in A_s \}$. Then, the proposed method is based on e-values, defined for each sensor $k$ at time $t$ as:
\begin{equation}\label{e:statistic}
    E_{k,t} = \frac{1}{t} \sum_{s=1}^{t} \prod_{\substack{s \leq i \leq t \\ i \in C_{k,t}}} L_{k,i}.
\end{equation}
Next, we arrange the e-values in descending order. Denote by $[|A_t|] = \{1,\cdots,|A_t|\}$ where $|A_t|$ is the size of the set $A_t$, and $E_{(k,t)}$ the $k$-th largest e-value among $(E_{k,t})_{k \in A_t}$. More precisely, there exists a mapping $\sigma: [|A_t|] \to A_t$ such that $\{\sigma(k): k \in [|A_t|]\} = A_t$ and $E_{(k,t)} = E_{\sigma(k),t}$ for all $k \in [|A_t|]$, with
$
E_{(1,t)} \geq E_{(2,t)} \geq \cdots \geq E_{(|A_t|,t)}.
$
We then define
\begin{equation} \label{e:numrej}
R_t = \max\left\{ k \in [|A_t|] : \frac{k\, E_{(k,t)}}{K} \geq \frac{1}{\alpha} \right\},
\end{equation}
where $K$ is the almost sure upper bound on $|A_t|$ defined in Assumption \ref{con:restriction}.
By convention, we let $R_t=0$ if $\{ k \in [|A_t|] : \frac{k\, E_{(k,t)}}{K} \geq \frac{1}{\alpha} \}=\emptyset$.
Finally, we declare a change in the sensors corresponding to the top $R_t$ e-values:
$$
D_{t+1} = 
\begin{cases}
\{\sigma(k) : k \in [R_t]\} & \text{if $R_t > 0$}
\\ \emptyset & \text{if $R_t = 0$}.
\end{cases}
$$
The proposed method is summarized in Algorithm~\ref{alg:alg1}, which we will refer to as the \underline{Sh}iryaev-Roberts \underline{E}-value-based Benjamini-Hochberg procedure for \underline{Re}configurable Sensors with \underline{F}DR control (SHEREF) method in later sections.
\begin{algorithm}[H]
\caption{{Shiryaev-Roberts E-value-based Benjamini-Hochberg procedure for Reconfigurable Sensors with FDR control (SHEREF).}} \label{alg:alg1}
\begin{algorithmic}[1]
\State \textbf{Input:} target level $\al$, running time $\bar{T}$.
\State Initialization: $t \gets 1$, $D_1 = \emptyset$.
\While {$1 \leq t \leq \bar{T}-1$ and $A_t \neq \emptyset$}
\State Compute $(E_{k,t})_{k \in A_t}$ according to \eqref{e:statistic}.
\State Order $(E_{k,t})_{k \in A_t}$ from largest to smallest:  $E_{(1,t)} \geq \cdots \geq E_{(|A_t|,t)}$.
\State $R_t \gets \max \{k \in [|A_t|] \mid \frac{k E_{(k,t)}}{K} \geq \frac{1}{\al} \}$.
\State $D_{t+1} \gets \text{set of sensor indices corresponding}$ to the largest $R_t$ values of $(E_{k,t})_{k \in A_t}$.
\State $t \gets t+1$.
\EndWhile
\State \textbf{Output:} Detection set $(D_t)_{1 \leq t \leq \bar{T}-1}$.
\end{algorithmic}
\end{algorithm}
The proposed method guarantees that $\text{maxFDR} \leq \alpha$ (see Section \ref{theoreticalresults} for rigorous justifications). Below, we explain the connection between the proposed SHEREF method and the e-Benjamini–Hochberg (e-BH) method in multiple testing \cite{vovk2021}.

First, for each $k \in A_t$, consider the hypothesis testing problem
\begin{equation}\label{eq:hypothesis}
    H_{0,k,t}: t \leq \tau_k \quad \text{versus} \quad H_{1,k,t}: t > \tau_k.
\end{equation}
Under $H_{0,k,t}$, sensor $k$ remains in the pre-change regime, whereas under $H_{1,k,t}$ it transitions into the post-change regime. The statistic $L_{k,i}$ defined in \eqref{eq:L} serves as a test statistic, where larger values indicate stronger evidence against $H_{0,k,i}$. To enhance detection efficiency, we aggregate the $L_{k,i}$ values over $1 \leq i \leq t$ and construct $E_{k,t}$ according to \eqref{e:statistic}. Notably, if $k \in A_i$ for all $1 \leq i \leq t$, then $tE_{k,t}$ coincides with the Shiryaev–Roberts statistic, which is a well-known minimax statistic for single-stream change detection \cite{polunchenko2010,shiryaev1963optimum,pollak1985optimal}. 

Second, $E_{k,t}$ is closely related to the notion of an e-value in the multiple testing literature \cite{vovk2021}. E-values are non-negative test statistics with an expected value of at most 1 under the null hypothesis, where larger values indicate more evidence against the null. {We use $\E_{H_{0,k,t}}[\cdot]$ to denote the expectation under the null hypothesis $H_{0,k,t}$, which indicates that the $k$-th sensor is in pre-change regime at time $t$, i.e., $t \leq \tau_k$.} In our context, we show that
$$
\Eb_{H_{0,k,t}}[E_{k,t}] \leq 1 \quad \text{for all } k \text{ and } t,
$$
which confirms that $E_{k,t}$ qualifies as an e-value for the testing problem in \eqref{eq:hypothesis}.
Thus, $E_{k,t}$ is not only well-suited for change detection but can also be seamlessly integrated into an e-value-based multiple-testing framework. Following the e-BH procedure developed in \cite{wang2022}, we then form the detection set $D_{t+1}$.
Third, we note that Assumption \ref{ass2} guarantees that both $L_{k,i}$ and $E_{k,t}$ qualify as test statistics, as they depend solely on the observed data and are independent of the unknown change points.

\subsection{Boosting Procedure} \label{boostingprocedure}
\noindent 
In this section, we propose methods to improve the power of SHEREF. That is, to reduce the false negative rate while maintaining control of the maximum false discovery rate ($\text{maxFDR}$) at the desired level. We introduce two methods for different dependence structures.

The first dependence structure, referred to as {\it general dependence (GD)}, allows the processes $X_{k,t}$ to follow arbitrary dependence, provided that they satisfy Assumptions \ref{ass2} and \ref{con:restriction}. The second structure, termed {\it time independent positive dependence (TIPD)}, assumes that the observations are independent across time while being positively dependent across sensors.  The precise definition is provided below.

\begin{assumption}[{Time independent positive dependence (TIPD)}] \label{def:tipd}
Given $A_t$, define the likelihood ratio for $k \in A_t$ as
\begin{equation} \label{e:lr}
\tilde{L}_{k,t} = \frac{f_{1,k,t}(X_{k,t} \mid \Xc_{t-1}, \Ab_{t})}{f_{0,k,t}(X_{k,t} \mid \Xc_{t-1}, \Ab_{t})}.
\end{equation}
For each $t$, define the random vector $\tilde{\Lb}_t = (\tilde{L}_{k,t})_{k \in A_t}$. 
Then, TIPD is satisfied if the following two conditions hold for all  $t$:

\begin{enumerate}
\item{{Time independence}: For all $k, t \in \Zb_{+}$, $X_{k,t}$ is independent of $\F_{t-1}$.} 
\item{{Positive dependence}:
For all $k \in A_t$ and all decreasing set $B \subset \RR^{K}$ the function: $x \rightarrow P(\tilde{\Lb}_t \in B \mid \tilde{L}_{k,t} \geq x, \Xc_{t-1}, \Ab_{t})$ is nonincreasing on $[0,\infty)$, where a decreasing set  $B \subset \RR^{|A_t|}$ is a set satisfying $x \in B$ implies $y \in B$ for all $y \leq x$ (elementwise).}
\end{enumerate}
\end{assumption}
The TIPD condition generalizes the positive regression dependence on subsets (PRDS) condition introduced in \cite{benjamini2001,finner2009} for multiple testing to the setting of a reconfigurable sensor network. 
In Section \ref{discuss2}, we provide examples of models that satisfy the TIPD assumption.

For both GD and TIPD, our boosting method is based on the construction of boosted e-values:
\begin{equation}
E_{k,t}^{b} = 1_{(k \in A_t)} \, b_{k,t}\, E_{k,t},
\end{equation}
where the boosting factor $b_{k,t}$ is determined differently depending on the assumed dependence structure. Specifically, define 
\begin{equation} \label{e:gdset}
B_1 = \Bigg\{ b \geq 1 \Big| \E_{H_{0,k,t}} \Bigg[b  E_{k,t} 1_{\left(\al b E_{k,t} \geq 1 \right)} \Big| \F_{t-1}, A_t \Bigg] \\ \leq  \E_{H_{0,k,t}}[E_{k,t} \mid \F_{t-1}, A_{t}] \Bigg\},
\end{equation}
\begin{equation}  \label{e:tipdset}
B_2 = \Bigg \{ b \geq 1 \Big| \underset{1 \leq y \leq K}{\max} \, y P_{H_{0,k,t}} \left( b E_{k,t} \geq \frac{y}{\al} \Big| \F_{t-1}, A_t \right) \\
\leq \al \, \E_{H_{0,k,t}}[E_{k,t} \mid \F_{t-1}, A_t]  \Bigg \}.
\end{equation}
According to Theorem \ref{th:conservative}, both $B_1$ and $B_2$ are non-empty. For GD, we define the boosting factor  $b_{k,t}$  as 
\begin{equation} \label{e:gd2}
b_{k,t} = 
\begin{cases}
\sup B_1 & \text{if} \, \sup B_1 \in B_1
\\ \sup B_1 - \de_1 & \text{if} \, \sup B_1 \notin B_1.
\end{cases}
\end{equation}
where $\de_1 > 0$ is a small positive number.
For TIPD, we define the boosting factor $b_{k,t}$ as 
\begin{equation} \label{e:tipd2}
b_{k,t} = 
\begin{cases}
\sup B_2 & \text{if} \, \sup B_2 \in B_2
\\ \sup B_2 - \de_2 & \text{if} \, \sup B_2 \notin B_2.
\end{cases}
\end{equation}
where $\de_2 > 0$ is a small positive number.
Boosting factors defined in \eqref{e:gd2} and \eqref{e:tipd2} extends the boosting methods introduced in \cite{wang2022} to account for a  dynamically changing sensor network. Ideally, we aim to choose the $b_{k,t}$ such that inequalities in $B_1$ and $B_2$ hold with equalities. However, such $b_{k,t}$ may not always exist when the null distribution of $E_{k,t}$ is not continuous. In such cases, we select the value of $b_{k,t}$ as large as possible among those that satisfy the inequalities. With the boosted e-values $E_{k,t}^{b}$, the boosted change detection algorithm is implemented by substituting $E_{k,t}^{b}$ for $E_{k,t}$ in Algorithm \ref{alg:alg1}. Throughout the remainder of the paper, we refer to the boosted change detection methods under GD and TIPD as SHEREF-GD and SHEREF-TIPD, respectively. Both SHEREF-GD and SHEREF-TIPD increase the number of detections while maintaining control of the $\text{maxFDR}$ at the pre-specified level. Rigorous theoretical justifications for these methods are provided in Section \ref{theoreticalresults}.

\section{Theoretical Results} \label{theoreticalresults}
\noindent 

In this section, we present theoretical results for the proposed methods. The following theorem establishes a key property of the statistic $E_{k,t}$.

\begin{theorem} \label{th:evalue} 
Under Assumptions \ref{ass2} and \ref{con:restriction}, for each time $t \geq 1$ and sensor $k$, the statistic $E_{k,t}$, as defined in (\ref{e:statistic}), satisfies 
\begin{equation}
E_{k,t} \geq 0, \quad \text{and} \quad \E_{H_{0,k,t}}[E_{k,t}] \leq 1. 
\end{equation} 
\end{theorem}

\begin{proof} See Appendix. \end{proof}

The next theorem provides upper bounds on $\text{FDR}_t$ and $\text{maxFDR}$ for the SHEREF algorithm.

\begin{theorem} \label{th:procedure} 
Under Assumptions \ref{ass2} and \ref{con:restriction}, Algorithm \ref{alg:alg1} satisfies the following bounds for all $t \geq 1$: 
\begin{equation}\label{eq:FDR-bounds} 
\text{FDR}_t \leq \alpha \quad \text{and} \quad \text{maxFDR} \leq \alpha. 
\end{equation} \end{theorem}

\begin{proof} See Appendix. \end{proof}
The subsequent theorem extends these FDR bounds to the boosted change detection methods, SHEREF-GD and SHEREF-TIPD.

\begin{theorem} \label{th:gdprocedure} 
Suppose Assumptions \ref{ass2} and \ref{con:restriction} hold. Then, the bounds in \eqref{eq:FDR-bounds} are satisfied when using SHEREF-GD. Moreover, if Assumption \ref{def:tipd} is also satisfied, the bounds in \eqref{eq:FDR-bounds} hold for SHEREF-TIPD. 
\end{theorem}

\begin{proof} See Appendix. \end{proof}
The next theorem shows the relationship among SHEREF, SHEREF-GD, and SHEREF-TIPD.
\begin{theorem} \label{th:conservative}
Suppose Assumptions \ref{ass2} and \ref{con:restriction} hold. Then, $B_1$ and $B_2$ are nonempty. Let $b_{k,t}^{GD}$ and $b_{k,t}^{TIPD}$ denote the boosting factors defined in \eqref{e:gd2} and \eqref{e:tipd2} respectively. Given the same realization value for ($\Xc_{t-1}, \Ab_t$), and given $\de_1=\de_2 \leq  \sup B_2 - \sup B_1$, for each time $t \geq 1$ and each sensor $k$, we obtain:
\begin{equation}
b_{k,t}^{GD} \in B_1 ,\quad b_{k,t}^{TIPD} \in B_2 ,\quad 1 \leq b_{k,t}^{GD} \leq b_{k,t}^{TIPD}.
\end{equation}
\end{theorem}

\begin{proof} See Appendix. \end{proof}

The result of Theorem \ref{th:conservative} implies the following order of detection statistics:
$$
1_{(k \in A_t)} E_{k,t} \leq 1_{(k \in A_t)} b_{k,t}^{GD} E_{k,t} \leq 1_{(k \in A_t)} b_{k,t}^{TIPD} E_{k,t}.
$$
Since Algorithm \ref{alg:alg1} prioritizes detecting changes based on the larger statistics, a greater boosting factor is likely to result in more detections. Furthermore, by combining Theorems \ref{th:procedure} and \ref{th:gdprocedure}, SHEREF-TIPD is expected to be the least conservative method maintaining $\text{FDR}_t$ control when Assumptions  \ref{ass2}, \ref{con:restriction} and \ref{def:tipd} hold.

\section{Examples} \label{discuss2}
\noindent 
In this section, we present examples of specific models that satisfy Assumptions \ref{ass2}, \ref{con:restriction}, and \ref{def:tipd}. 
We start with the independent case.
\begin{example}[Independent observations] \label{ex:indep} \hfill{} 
Consider the following generation mechanism of data:
\begin{enumerate} \label{con:indep1}
\item $X_{k,t}\in \mathbb{R}^d$ are generated independently for different $k$ and $t \in \Zb_{+}$ and $\Ab_{t-1}$. In addition, $X_{k,t}$ has the marginal probability density functions $f_{0,k,t}(x)$ and $f_{1,k,t}(x)$ for $x\in \mathbb{R}^{d}$ and $t \leq \tau_k$ and $t > \tau_k$ respectively. Additionally, the support of $f_{1,k,t}$ is a subset of that of $f_{0,k,t}$.
\item $\Ab_t$ is externally controlled, and it satisfies Assumption \ref{con:restriction}.
\end{enumerate}
Then, following Theorem \ref{th:validex1} at the end of the section, Assumption \ref{ass2} is met under above data generation mechanism, and Assumption \ref{def:tipd} is met if the following statement also holds:
\begin{enumerate} \label{con:indep2}
\item{Assume $d=1$, and for each $k,t \in \Zb_{+}$, there exists a function $\xi_{k,t}$ such that $f_{1,k,t}(x)/f_{0,k,t}(x) = \xi_{k,t}(x)$. Also, $\xi_{k,t}$ is either nondecreasing or nonincreasing in $x$ for all $k \in \Zb_{+}$}. 
\end{enumerate}  
The above statement suggests that the assumptions can be greatly simplified under the independent case. This data generation process aligns with that introduced in \cite{chen2023compound, lu2022, chen2016}. However, our proposed procedure extends beyond the existing framework by accommodating dynamically changing active set $A_t$ under a reconfigurable network.

For a more concrete example satisfying all the above assumptions, we consider a mean-change model assuming $d = 1$:
\begin{equation} \label{e:indep}
X_{k,t} = 
\begin{cases}
\ep_{k,t} & t \leq \tau_k
\\ \mu + \ep_{k,t} & t > \tau_k.
\end{cases}
\end{equation}
Here, $\mu$ is a real constant, the collection of $\{\ep_{k,t}\}_{k \in \Zb_{+}}$ are jointly independent, following $N(0,\sigma^2)$ for $\sigma^2>0$. 
\end{example}

Next, we consider the case of concurrent dependence among sensors while maintaining independence across time. 

\begin{example}[Independence across time and concurrent dependence over sensors] \label{ex:condep} \hfill{}
\\Consider the following generation mechanism of data:
\begin{enumerate}
    \item For some $d_1, d_2$, $Z_{t}\in \mathbb{R}^{d_1}$ and $\epsilon_{k,t}\in \mathbb{R}^{d_2}$ are generated independently for different $t$, and $\Ab_{t-1}$. In addition, $\{Z_{t}\}_{t\in\Zb_+}$ and $\{\ep_{k,t}\}_{k,t\in\Zb_+}$ are independent. 
    \item For each $k,t \in \Zb_{+}$, there exist functions $h_{l,k,t}:\mathbb{R}^{d_1} \to \mathbb{R}^{d_2}$ for $l=0,1$, such that
    $X_{k,t}=h_{0,k,t}(Z_t)+\epsilon_{k,t}$ for $t\leq \tau_k$ and $X_{k,t}=h_{1,k,t}(Z_t)+\epsilon_{k,t}$ for $t> \tau_k$.
    \item For each $k,t \in \Zb_{+}$, $Z_t$ has the marginal density function $f_{Z_t}(z)$ with respect to the Lebesgue measure, and $\ep_{k,t}$ have marginal density function $f_{\ep_{k,t}}(e)$. Moreover, $f_{\ep_{k,t}}(e)$ has the support $\mathbb{R}^{d_2}$.
    \item $\Ab_t$ is selected satisfying Assumption \ref{con:restriction}.
\end{enumerate}
Then, following Theorem \ref{th:validex1}, Assumption \ref{ass2} is met under the above data generation mechanism, and Assumption \ref{def:tipd} is met if the $\{X_{k,t}\}$ is generated as follows.
\begin{enumerate}
\item{Let $H = \Zb_{+} \times \Zb_{+}$, and $\{X_{k,t}\}_{(k,t)\in H}$ be a Gaussian random field  satisfying
\begin{equation} \label{e:meanft}
E(X_{k,t}) = 
\begin{cases}
\de_0 & t \leq \tau_k
\\ \de_1 & t > \tau_k,
\end{cases}
\end{equation}
for some $\de_0, \de_1 \in \mathbb{R}$.
Additionally, the covariance satisfies $Cov(X_{k_1,t},X_{k_2,t})\geq 0$ for all $t$, and $Cov(X_{k_1,t_1},X_{k_2,t_2})=0$ if $t_1\neq t_2$.
}
\end{enumerate}

For a more concrete example, we can consider the following model for each $t$ and $k$ and $d_2 =1$:
\begin{equation} \label{e:condep}
X_{k,t} = 
\begin{cases}
a Z_t + \ep_{k,t} & t \leq \tau_k
\\ a Z_t + \mu + \ep_{k,t} & t > \tau_k,
\end{cases}
\end{equation}
where $Z_t$ and $\epsilon_{k,t}$ independently follows $N(0,\sig^2)$,  and $a\neq 0$ is a real constant. 
This structure captures scenarios where simultaneous observations are correlated due to an underlying shared factor.
\end{example}

We now consider models with temporal dependence. We explore iterative data generation processes, where current observations can be written as a function of past observations plus time-independent noises. 

\begin{example}[Dependence within sensors and across time] \label{ex:depstrm} \hfill{}
\\Consider the following data generation:
\begin{enumerate}
\item{For some $d_1$ and $d_2$, $Y_{k,t-1} \in \mathbb{R}^{d_1}$ is generated such that it is $\F_{t-1}$ measurable. Additionally, $\ep_{k,t} \in \mathbb{R}^{d_2}$ are generated independently for different $t$. Moreover, for all $t \in \Zb_{+}$, $\{\ep_{k,t}\}_{k \in \Zb_{+}}$ is independent of $\F_{t-1}$.}
\item{For each $k,t \in \Zb_{+}$, there exist functions $h_{l,k,t}: \mathbb{R}^{d_1} \to \mathbb{R}^{d_2}$ for $l = 0,1$, such that $X_{k,t} = h_{0,k,t}(Y_{k,t-1}) + \ep_{k,t}$ for $t \leq \tau_k$, and $X_{k,t} = h_{1,k,t}(Y_{k,t-1}) + \ep_{k,t}$ for $t > \tau_k$.}
\item{For each $k,t \in \Zb_{+}$, $\ep_{k,t}$ has the marginal density function $f_{\ep_{k,t}}(e)$  with the support equal to $\mathbb{R}^{d_2}$.}
\item{$\Ab_t$ is selected satisfying Assumption \ref{con:restriction}.}
\end{enumerate}
Then, from Theorem \ref{th:validex1}, Assumption \ref{ass2} holds under the above data generation mechanism.
For a more concrete example, consider the following model for each time $t$ and stream $k$, with $d_2=1$:
\begin{equation} \label{e:depwithstm}
X_{k,t} = 
\begin{cases}
a_{k,0} 1_{(k \in A_{t-1})} \, X_{k,t-1} + \ep_{k,t} & t \leq \tau_k
\\ a_{k,1} 1_{(k \in A_{t-1})} \, X_{k,t-1} + \ep_{k,t} & t > \tau_k.
\end{cases}
\end{equation}
Here, $a_{k,0}$ and $a_{k,1}$ are stream-specific constant, satisfying $|a_{k,i}| < 1$ for $i \in \{0,1\}$. Also, any subset of $\{\ep_{k,t}\}_{k \in \Zb_{+}}$ is a mean-zero multivariate Gaussian vector independent of $\F_{t-1}$. 
\end{example}

Next, we consider the scenario where $A_t = [K]$ for all $t$, meaning that all sensors remain active at every time point. In applications such as spectrum sensing, this setup corresponds to continuously receiving sensor data even after detecting changes in some of the sensors.

\begin{example}[Dependent across time and sensors with fixed active set] \label{ex:fixindex} \hfill{}
\\ Consider the following data generation:
\begin{enumerate}
\item{$A_t = [K]$ for all $t \in \Zb_{+}$.}
\item{For some $d$, $X_{k,t} \in \mathbb{R}^{d}$ has conditional probability density functions $f_{0,k,t}(x \mid (x_{k,s})_{k \in [K], 1 \leq s \leq t-1})$ and $f_{1,k,t}(x \mid (x_{k,s})_{k \in [K], 1 \leq s \leq t-1})$ for $x \in \mathbb{R}^d$ and $t \leq \tau_k$ and $t > \tau_k$ respectively. Moreover, the support of $f_{1,k,t}$ is a subset of the support of $f_{0,k,t}$.} 
\end{enumerate}
Then, following Theorem \ref{th:validex1}, Assumptions \ref{ass2} and \ref{con:restriction} are satisfied under the above data generation mechanism.
For example, we can consider the following model for each time $t$ and sensor $k$ with $d=1$:
\begin{equation} \label{e:dep}
X_{k,t} = 
\begin{cases}
\sum_{l = 1}^{K} a_l X_{l,t-1} + \ep_{k,t} & t \leq \tau_k
\\ \sum_{l = 1}^{K} a_l X_{l,t-1} + \mu + \ep_{k,t} & t > \tau_k,
\end{cases}
\end{equation}
where $\mu$ is real constant, $a_l$ is constant satisfying $|a_l| < 1$, and for each $t$, and $\{\ep_{k,t}\}_{k \in \Zb_{+}}$ is a mean-zero Gaussian random field. This example demonstrates that the framework allows a very general dependence structure when the sensor network remains fixed over time.
\end{example}

\begin{theorem} \label{th:validex1}
For examples \ref{ex:indep}, \ref{ex:condep}, \ref{ex:depstrm}, \ref{ex:fixindex}, the following statements hold.
\begin{enumerate}
\item{Assumption \ref{ass2} is satisfied in Examples \ref{ex:indep}, \ref{ex:condep}, and \ref{ex:depstrm}. For Example \ref{ex:fixindex}, both Assumptions \ref{ass2} and \ref{con:restriction} hold if the respective conditions in each example are met.}
\item{Assumption \ref{def:tipd} is satisfied in examples \ref{ex:indep} and \ref{ex:condep} when given conditions in each example hold.}
\end{enumerate}
\end{theorem}

\begin{proof} See Appendix. \end{proof}

\section{Numerical Studies} \label{Simulationstudies}
\noindent We present the empirical performance evaluation of the SHEREF and its boosted versions SHEREF-GD and SHEREF-TIPD. The performance of the proposed methods is compared with two alternative methods: anytime valid p-value (AVP) \cite{johari2015} and D-FDR \cite{chen2016}. 
We note that the AVP method was initially designed for multiple hypothesis testing. To apply it to a change detection problem, we first reformulate the change detection problem as a multiple testing problem, as shown in \eqref{eq:hypothesis}. When the observations $X_{k,t}$ are independent across time and streams, D-FDR is theoretically justified in controlling $\text{maxFDR}$ at a desired level under the assumption that $A_t = A_{t-1}\setminus D_t$, whereas AVP is theoretically justified in controlling $\text{maxFDR}$ under the assumption that $A_t = [K]$ for all $t$. However, unlike the proposed methods, AVP and D-FDR lack theoretical support for a dynamically changing sensor network.

For the simulation study, we set $\bar{T}=30$, and $|A_t| = 100$ for all $1 \leq t \leq \bar{T}$. This is achieved by considering a larger pool of $1000$ streams and using the first $100$ streams at the beginning of the detection procedure. We then deactivate the detected sensors for the remaining time points and randomly replace all of the detected sensors with the streams that have not been used in the detection procedure. Additionally, the change point $\tau_k$ for each sensor $k$ is randomly generated from a geometric distribution with $P(\tau_k = t) = (0.9)^t (0.1) $ for $t\in\Zb_+\cup \{0\}$.
For generating the observations, we consider the following two models:
\begin{itemize}
    \item Model 1: 
    \begin{equation} \label{e:simulation}
X_{k,t} = 
\begin{cases}
a_k Z_t + \ep_{k,t} & t \leq \tau_k
\\ a_k Z_t + 3 + \ep_{k,t} & t > \tau_k,
\end{cases}
\end{equation}
 for $k\in\Zb_+$ and time $1 \leq t \leq \bar{T}$, 
where $Z_t$ independently follows $N(0,1)$, $a_k$ is randomly sampled from $\text{Unif}(0,0.5)$, and $\ep_{k,t}$ follows an independent standard normal distribution for each $t$. This model allows for dependence across time and independence over sensors, satisfying the Assumption \ref{def:tipd} as mentioned in Example \ref{ex:condep}.
\item Model 2:
\begin{equation} \label{e:simulation2}
X_{k,t} =  
\begin{cases}
-0.8 \, (1_{(k \in A_{t-1})} X_{k,t-1}) + \ep_{k,t} & t \leq \tau_k
\\ 0.8 \, (1_{(k \in A_{t-1})} X_{k,t-1})  + \ep_{k,t} & t > \tau_k,
\end{cases}
\end{equation}
where, for each $t$, let $\tilde{\ep}_{t}$ denote the vector of $(\ep_{1,t},\cdots,\ep_{1000,t})$. The sequence $(\tilde{\ep}_t)_{t \geq 1}$ is generated such that it is independent across time and for each fixed $t$, $\tilde{\ep}_{t}$ follows a multivariate Gaussian distribution $N(0, \Sig)$. The covariance matrix $\Sig$ satisfies $\Sig_{i,j} = (-0.8)^{|i-j|}$. The model allows for dependence both across time and sensors,  as mentioned in Example \ref{ex:depstrm}. We note that under this model, SHEREF and SHEREF-GD are theoretically guaranteed to control the $\text{maxFDR}$, whereas other methods, including SHEREF-TIPD, AVP, and D-FDR, do not have theoretical justification.
\end{itemize}
We evaluate performance using the \text{AFNR} defined in \eqref{e:afnr}, along with an additional metric called total average detection delay (\text{TADD}), defined as:
\begin{equation}\label{e:tadd}
\text{TADD} = \E \left[ \sum_{t=1}^{\bar{T}-1} \sum_{k \in A_t} 1_{(\tau_k < t,\; k \notin D_{t+1})} \right].
\end{equation}
\text{TADD} quantifies the total number of false negatives accumulated until a predetermined constant $\bar{T}$.

The performance of the five different procedures is evaluated based on the estimated values of \text{AFNR}, \text{TADD}, \text{maxFDR}, and the total number of detections. We perform Monte Carlo simulations with $500$ replications for two target levels, $\al=0.1$ and $\al=0.05$.
The simulation results are summarized in Tables \ref{t:addsimulation} and \ref{t:addsimulation2}.
For each metric, the value on the left corresponds to $\al = 0.1$, while the value on the right corresponds to $\al =0.05$. 
Despite the lack of theoretical guarantees of AVP and D-FDR, they control \text{maxFDR} at the target level in the simulation, probably because these methods are conservative in making detections.

Table \ref{t:addsimulation} and Table \ref{t:addsimulation2} show that
SHEREF and its boosted variations consistently outperform the competing methods in the numerical study, as suggested by their lower values in \text{AFNR} and \text{TADD}. 
Also, the effectiveness of the boosting method is demonstrated with improved performance across all evaluation metrics. In particular, SHEREF-TIPD achieves the lowest \text{AFNR} and \text{TADD}, which aligns with the Theorem \ref{th:conservative}. 

\begin{table}[!htp]
\centering
\caption{Simulation results for model (\ref{e:simulation}).}
\label{t:addsimulation}
\begin{tabular}{cccccc}
\toprule
$\al=0.1/0.05$ &AFNR &TADD &max($\text{FDR}_t$) &Number of rejections \\
\midrule
SHEREF &0.2114/0.2444 &466.404/546.706 &0.0125/0.0049 &794.592/763.732 \\
\midrule
SHEREF-GD &0.1951/0.2253 &427.256/499.972 &0.0146/0.0089 &810.47/781.694 \\
\midrule
SHEREF-TIPD &0.1133/0.1453 &237.982/310.806 &0.0813/0.0448 &901.168/861.772 \\
\midrule
AVP & 0.5289/0.5430 &1367.68/1416.042 & 0.0123/0.0056 & 415.014/393.034 \\
\midrule
D-FDR & 0.3509/0.3629 &825.706/857.096 & 0.0041/0.0023 & 647.53/635.52 \\
\bottomrule
\end{tabular}
\end{table}

\begin{table}[!htp]\centering
\caption{Simulation results for model (\ref{e:simulation2})}
\label{t:addsimulation2}
\begin{tabular}{cccccc}
\toprule
$\alpha$=0.1/0.05 &AFNR &TADD &max($\text{FDR}_t$) &Number of rejections \\
\midrule
SHEREF &0.4540/0.4735 &1078.142/1137.522 &0.0038/0.0023 &625.742/598.058 \\
\midrule
SHEREF-GD &0.4424/0.4630 &1041.49/1103.976 &0.0046/0.0032 &648.826/616.546 \\
\midrule
SHEREF-TIPD &0.3927/0.4194 &895.55/974.742 &0.0227/0.0107 &719.958/676.676 \\
\midrule
AVP &0.5770/0.5823 &1540.374/1558.234 &0.0044/0.0020 &330.656/324.478 \\
\midrule
D-FDR &0.5426/0.5506 &1368.942/1395.87 &0.0002/0.00009 &647.53/635.52 \\
\bottomrule
\end{tabular}
\end{table}

\section{Conclusion} \label{conclusion}
\noindent In this paper, we proposed change detection methods to control the FDR for reconfigurable sensor networks. The proposed framework accommodates general dependence structures and dynamically changing sensor configurations, which is substantially more general than the settings of the existing works. Numerical experiments demonstrate that,
with the same false discovery rate target, our procedures achieve
superior performance compared to existing methods, exhibiting
lower false non-discovery rates and reduced detection delays. 

The current research inspires several promising future directions. First, the boosting method SHEREF-TIPD always outperforms the rest in our numerical experiments, even when the TIPD assumption is not satisfied. As the TIPD assumption is relatively strong, it is interesting to investigate this method further to see if the resulting FDR can be controlled under weaker assumptions.
Second, in practice, the pre- and post-change distributions may not be fully known or accurately modeled. Therefore, it is worth extending the current procedure to settings where these distributions are unknown. 

\appendices 
\section{Proof of Theoretical Results in Sections \ref{methods}, \ref{theoreticalresults}, and \ref{discuss2}} \label{appendA}
\noindent Throughout the proofs, we write 
$$
f_{i,k,t}(X_{k,t} \mid \Xc_{t-1}, \Ab_{t}) = f_{i,k,t}(X_{k,t} \mid \F_{t-1}, A_t),
$$ 
for all $i \in \{0,1\}$ and for each $k$ and $t$.

\begin{proof}[Proof of Theorem \ref{th:evalue}]
The non-negativity follows since all probability densities are non-negative, and for all $k, t \in \Zb_{+}$, $L_{k,t}$ are well defined from Assumption \ref{ass2}. We now prove the second statement. For given $A_t$, define $\tilde{L}_{k,t}$ for $k \in A_t$ as
\begin{equation*}
   \tilde{L}_{k,t} = \frac{f_{1,k,t}(X_{k,t} \mid \F_{t-1}, A_t)}{f_{0,k,t}(X_{k,t} \mid \F_{t-1}, A_t)}.
\end{equation*}
According to Assumption \ref{con:restriction}, the conditional distribution of $X_{k,t}|\F_{t-1},A_t$ does not depend on $A_t$. Thus, the conditional density functions $f_{i,k,t}(X_{k,t} \mid \F_{t-1}, A_t) $ does not depend on $A_{t}$ and can be written as $f_{i,k,t}(X_{k,t} \mid \F_{t-1})$ for $i=0,1$. 
That is,
\begin{equation} \label{e:condenft}
f_{i,k,t}(X_{k,t} \mid \F_{t-1}, A_t) = f_{i,k,t}(X_{k,t} \mid \F_{t-1}) \quad \text{a.s}. 
\end{equation}
From \eqref{e:condenft}, $\tilde{L}_{k,t} = \frac{f_{1,k,t}(X_{k,t} \mid \F_{t-1})}{f_{0,k,t}(X_{k,t} \mid \F_{t-1})}$ almost surely. Now, we prove that for any $k \in \Zb_{+}$ and $t \geq 1$,
\begin{equation} \label{e:thm1pf1}
\E_{H_{0,k,t}}[1_{(k \in A_t)}\tilde{L}_{k,t} \mid \F_{t-1}] = \E_{H_{0,k,t}}[1_{(k \in A_t)} \mid \F_{t-1}] \quad \text{a.s}.
\end{equation}
To see this, applying the law of iterated expectations
\begin{align*}
\E_{H_{0,k,t}}[1_{(k \in A_t)} \tilde{L}_{k,t} \mid \F_{t-1}] 
&= \E_{H_{0,k,t}}[ \E_{H_{0,k,t}}[1_{(k \in A_t)}\tilde{L}_{k,t} \mid \F_{t-1}, A_t] \mid \F_{t-1}]
\\ &= \E_{H_{0,k,t}}[1_{(k \in A_t)} \E_{H_{0,k,t}}[\tilde{L}_{k,t} \mid \F_{t-1}, A_t] \mid \F_{t-1}]
\\ &= \E_{H_{0,k,t}} \left[1_{(k \in A_t)} \E_{H_{0,k,t}} \left[\frac{f_{1,k,t}(X_{k,t} \mid \F_{t-1})}{f_{0,k,t}(X_{k,t} \mid \F_{t-1})} \Big | \F_{t-1} \right] \Big | \F_{t-1} \right]
\\ &= \E_{H_{0,k,t}}[1_{(k \in A_t)} \mid \F_{t-1}],
\end{align*}
where the third equality follows from \eqref{e:condenft}, and independent stream addition in Assumption \ref{con:restriction} which ensures $X_{k,t}$ is independent of $A_t$ given $\F_{t-1}$ for all $k,t \in \Zb_{+}$. 

Let 
\begin{equation*}
S_{k,t} = t E_{k,t}.
\end{equation*}
It satisfies the iterative updating equation
\begin{equation} \label{e:iterative}
S_{k,t} = 1_{(k \in A_t)} L_{k,t} (1+S_{k,t-1}) + 1_{(k \notin A_t)} S_{k,t-1}.
\end{equation}
Taking conditional expectation under $H_{0,k,t}$, we get
\begin{align*}
\E_{H_{0,k,t}}[ S_{k,t} \mid \F_{t-1}] = & (1 + S_{k,t-1}) \, \E_{H_{0,k,t}}[1_{(k \in A_t)} \tilde{L}_{k,t} \mid \F_{t-1}] + S_{k,t-1} \, \E_{H_{0,k,t}}[ 1_{(k \notin A_t)} \mid \F_{t-1}]
\\ =& (1 + S_{k,t-1}) \, \E_{H_{0,k,t}}[ 1_{(k \in A_t)} \mid \F_{t-1} ] + S_{k,t-1} \, \E_{H_{0,k,t}}[ 1_{(k \notin A_t)} \mid \F_{t-1}]
\\=& S_{k,t-1} + P_{H_{0,k,t}}(k \in A_t \mid F_{t-1}).
\end{align*}
The second equality follows from \eqref{e:thm1pf1}. Applying the expectation iteratively, we obtain the identity:
\begin{equation} \label{e:evalequal}
\E_{H_{0,k,t}}[S_{k,t}] = \sum_{s=1}^{t} P_{H_{0,k,t}}(k \in A_s). 
\end{equation}
We justify this as follows:
\begin{align*} 
\E_{H_{0,k,t}}[S_{k,t}] & = \E_{H_{0,k,t}}[ \E_{H_{0,k,t}}[ S_{k,t} \mid \F_{t-1}] ] 
\\ &= \E_{H_{0,k,t}}[ S_{k,t-1} ] + P_{H_{0,k,t}}(k \in A_t) 
\\ &= \E_{H_{0,k,t}}[ \E_{H_{0,k,t}}[S_{k,t-1} \mid \F_{t-2}] ] + P_{H_{0,k,t}}(k \in A_t) 
\\ &= \E_{H_{0,k,t}}[ S_{k,t-2} ] + P_{H_{0,k,t}}(k \in A_{t-1}) + P_{H_{0,k,t}}(k \in A_t) 
\\ & \vdots 
\\ &= \sum_{s=1}^{t} P_{H_{0,k,t}}( k \in A_s ).
\end{align*}
Since $P_{H_{0,k,t}}(k \in A_s) \leq 1$ for all $s \geq 1$, it follows that $\E_{H_{0,k,t}}[S_{k,t}] \leq t$. Finally, the statement follows from $\E_{H_{0,k,t}}[E_{k,t}] = \frac{1}{t} \E_{H_{0,k,t}}[S_{k,t}] \leq 1$.

\end{proof}

We note that the condition $\E_{H_{0,k,t}}[E_{k,t}] \leq 1$ is not sufficient to guarantee $\text{maxFDR} \leq \al$. A more delicate analysis is required, which we present in the next theorem.

\begin{proof}[Proof of Theorem \ref{th:procedure}]
Define $N_{0,t}$ as the set of indices corresponding to sensors under the pre-change regime:
\begin{equation*}
N_{0,t} = \{ k \in \Zb_{+}: t \leq \tau_k \}. 
\end{equation*}
From Assumption \ref{con:restriction}, we have the boundedness condition $\underset{t \geq 1}{\sup}|A_t| \leq K$. Consequently, for any $1 \leq r \leq t$, the number of active pre-change sensors is upper-bounded almost surely:
\begin{equation}
\sum_{k \in N_{0,t}} 1_{(k \in A_r)} \leq |A_r| \leq K.
\end{equation}
Taking expectations on both sides, 
we obtain:
\begin{equation} \label{e:bd1}
\sum_{k \in N_{0,t}} P_{H_{0,k,t}} ( k \in A_r ) = \E_{H_{0,k,t}}\Big[ \sum_{k \in N_{0,t}} 1_{(k \in A_r)}\Big] \leq K.
\end{equation}
Taking supremum over all $1 \leq r \leq t$, we obtain: 
\begin{equation} \label{e:bd2}
\underset{1 \leq r \leq t}{\sup} \underset{k \in N_{0,t}}{\sum} P_{H_{0,k,t}}(k \in A_r) \leq K.  
\end{equation} 
Combining this with \eqref{e:evalequal}, we derive the following bound:
\begin{equation} \label{e:efbd}
\sum_{k \in N_{0,t}} \E_{H_{0,k,t}}[E_{k,t}] \leq K.
\end{equation}
This result is obtained through the following derivation:
\begin{align*} 
\sum_{k \in N_{0,t}} \E_{H_{0,k,t}}[E_{k,t}] &= \frac{1}{t} \sum_{k \in N_{0,t}} \E_{H_{0,k,t}}[S_{k,t}] 
\\& = \frac{1}{t} \sum_{k \in N_{0,t}} \sum_{s=1}^{t} P_{H_{0,k,t}}(k \in A_s) 
\\& \leq \frac{1}{t} \sum_{s=1}^{t} \Big[ \underset{1 \leq r \leq t}{\sup} \underset{k \in N_{0,t}}{\sum} P_{H_{0,k,t}}(k \in A_r) \Big] 
\\ & \leq K.
\end{align*}
Recall that $R_t$ represents the number of detected sensors at time $t$, as defined in \eqref{e:numrej}, and let $V_t$ denote the number of false detections at time $t$. According to the proposed detection method, a change-point is detected in sensor $k$ at time $t$ is equivalent to the existence of $k'$ such that
$E_{(k',t)}=E_{k,t}$, $E_{(k',t)}\geq \frac{K}{k'\alpha}$ and $1\leq k'\leq R_t$. This implies that $E_{(k',t)}\geq \frac{K}{k'\alpha} \geq \frac{K}{R_t\alpha}$, which further implies 
\begin{equation} \label{e:FDRrej}
1_{(k \in A_t)} \, E_{k,t} \geq \frac{K}{\al R_t}.
\end{equation}
Then, the false discovery rate at time $t$ satisfies:
\begin{align*}
\text{FDR}_t & = \E \left[ \frac{V_t}{R_t \vee 1} \right]
\\ &= \sum_{k \in N_{0,t}} \E \left[ \frac{1_{(1_{(k \in A_t)} E_{k,t} \geq \frac{K}{\al R_t})}}{R_t \vee 1} \right]
\\ & \leq \frac{\al}{K} \sum_{k \in N_{0,t}} \E_{H_{0,k,t}} \left[ 1_{(k \in A_t)} E_{k,t} \right]
\\ & \leq \frac{\al}{K} \sum_{k \in N_{0,t}} \E_{H_{0,k,t}} \left[ E_{k,t} \right]
\\ & \leq \al.
\end{align*}
The second equality follows from \eqref{e:FDRrej}. The last inequality follows from \eqref{e:efbd}.
This proves the first statement. The second statement follows by taking supremum of $\text{FDR}_t$ over all values of $t \geq 1$. 
\end{proof}

\begin{proof}[Proof of Theorem \ref{th:gdprocedure}]

For each $t \geq 1$, define: 
\begin{equation*}
Q_t(u) = |\{ k \in \Zb_{+} \mid E_{k,t}^{b} \geq u \}| \vee 1,
\end{equation*}
and the detection threshold:
\begin{equation*}
u_{\al, t} = \inf \Big\{ u \in [0, \infty) \mid u Q_t(u) \geq \frac{K}{\al} \Big\}.
\end{equation*}
We first prove the theorem for SHEREF-GD. 
According to Lemma \ref{pf:lemma0} in Appendix \ref{appendB}, 
for each $t$, 
Algorithm \ref{alg:alg1} applied to $E_{k,t}^{b}$ under the target level $\al$ detects a change in the $k$-th sensor if and only if
\begin{equation*}
E_{k,t}^{b} \geq u_{\al,t},
\end{equation*}
and $u_{\al,t} Q_t(u_{\al,t}) = \frac{K}{\al}$. Since \eqref{e:FDRrej} also hold when replacing $E_{k,t}$ with $E_{k,t}^{b}$, we have $R_t = Q_t(u_{\al,t})$. %

We can verify the following for all $t \geq 1$:
\begin{align*}
\text{FDR}_t &= \E \left[ \frac{V_t}{R_t \vee 1} \right]
\\ &= \sum_{k \in N_{0,t}} \E \left[ \frac{1_{(E_{k,t}^{b} \geq u_{\al,t})}}{Q_t(u_{\al,t})} \right]
\\ &\leq \sum_{k \in N_{0,t}} \E \left[ \frac{1_{(b_{k,t}E_{k,t} \geq u_{\al,t})}}{Q_t(u_{\al,t})} \right] 
\\ &= \frac{\al}{K} \sum_{k \in N_{0,t}} \E \left[ u_{\al,t} 1_{(b_{k,t}E_{k,t} \geq u_{\al,t})}\right] 
\\ & = \frac{1}{K} \sum_{k \in N_{0,t}} \E \left[ \al u_{\al,t} 1_{(\al b_{k,t}E_{k,t} \geq \al u_{\al,t})}\right].
\end{align*}
Here, the first inequality is obtained by removing the indicator $1_{(k \in A_t)}$. The fourth line follows from the identity $u_{\al,t}Q_t(u_{\al,t}) = \frac{K}{\al}$. Using the lower bound $\al u_{\al,t} \geq 1$, we have the following upper bound when $\al b_{k,t} E_{k,t} \geq \al u_{\al,t}$:
\begin{align*}
\text{FDR}_t 
& \leq \frac{1}{K} \sum_{k \in N_{0,t}} \E_{H_{0,k,t}} \left[ \E_{H_{0,k,t}}\left[ \al b_{k,t}E_{k,t} 1_{(\al b_{k,t} E_{k,t} \geq 1)} \mid \F_{t-1}, A_t \right] \right]
\\ & \leq \frac{1}{K} \sum_{k \in N_{0,t}} \al \E_{H_{0,k,t}} \left[ \E_{H_{0,k,t}} \left[ E_{k,t} \mid \F_{t-1}, A_t \right] \right]
\\ & = \frac{\al}{K} \sum_{k \in N_{0,t}} \E_{H_{0,k,t}} [E_{k,t}]
\\ & \leq \al.
\end{align*}
The second inequality follows from the condition on $b_{k,t}$ in \eqref{e:gdset}, with $b_{k,t} \in B_1$, verified in Theorem \ref{th:conservative}. The bound on $\text{maxFDR}$ follows by taking supremum of $\text{FDR}_t$ over all values of $t \geq 1$..

Now we prove the result for SHEREF-TIPD. By definition, $u_{\al,t}$ is a function of $(E_{k,t}^b)_{k \in \Zb_+}$. Recall that $E_{k,t}^{b} = 1_{(k \in A_t)}b_{k,t}E_{k,t}$, where $b_{k,t}$ is $\F_{t-1}$ measurable and from \eqref{e:iterative}, $E_{k,t}$ is measurable with respect to $\sig((S_{k,t-1})_{k \in \Zb_+}, A_{t}, \tilde{L}_{k,t})$. Therefore, there exists a measurable function $\phi_t$ with respect to $\F_t$ such that $u_{\al,t} = \phi_t(\Qb_{t-1}, (\tilde{L}_{k,t})_{k \in A_t})$, where $\Qb_{t-1} = ((S_{k,t-1})_{k \in \Zb_+}, A_{t})$. Define $(E_{k,t}^{b})^{'}= \sup\{y \in I_{\phi_t} \cup \{0\} \mid y \leq E_{k,t}^{b}\}$, where $I_{\phi_t} = \frac{K}{\al [K]} = \{ \frac{K}{\al s} \mid s \in \Zb_{+}, 1 \leq s \leq K \}$ is the range of the measurable function $\phi_t$. Given $Q_t(u_{\al,t}) = R_t \vee 1$ and $u_{\al,t} Q_t(u_{\al,t}) = \frac{K}{\al}$, we derive the following bound on $\text{FDR}_t$:
\begin{align*}
\text{FDR}_t &= \E \left[ \frac{V_t}{R_t \vee 1} \right]
\\ &= \sum_{k \in N_{0,t}} \E \left[ \frac{1_{(E_{k,t}^{b} \geq u_{\al,t})}}{Q_t(u_{\al,t})} \right]
\\ &= \frac{\al}{K} \sum_{k \in N_{0,t}} \E \left[ u_{\al,t} 1_{(E_{k,t}^{b} \geq u_{\al,t})}\right]
\\ & = \frac{\al}{K} \sum_{k \in N_{0,t}} \E \left[ \E\left[u_{\al,t} 1_{(E_{k,t}^{b} \geq u_{\al,t})} \mid \F_{t-1}, A_t \right] \right].
\end{align*}
By construction, $u_{\al,t}$ is nonincreasing in $E_{k,t}^{b}$, and $E_{k,t}^{b}$ is nondecreasing in each component of $(\tilde{L}_{k,t})_{k \in A_t}$. Therefore, $u_{\al,t}$ is nonincreasing in each component of $(\tilde{L}_{k,t})_{k \in A_t}$. Applying Lemma \ref{pf:lemma2}, we obtain: 
\begin{align*}
\text{FDR}_t  
&  \leq \frac{\al}{K} \sum_{k \in N_{0,t}} \E \left[ \sup_{x \geq 0} x P( (E_{k,t}^{b})^{'} \geq x \mid \F_{t-1}, A_t ) \right]
\\ &  \leq \frac{\al}{K} \sum_{k \in N_{0,t}} \E \left[ \sup_{x \geq 0} x P( \al (b_{k,t} E_{k,t})^{'} \geq \al x \mid \F_{t-1}, A_t ) \right]
\\ &  = \frac{\al}{K} \sum_{k \in N_{0,t}} \E \left[ \sup_{x \geq 0} \frac{x}{\al} P( \al (b_{k,t} E_{k,t})^{'} \geq x \mid \F_{t-1}, A_t ) \right]
\\ &  = \frac{1}{K} \sum_{k \in N_{0,t}}  \E \left[ \sup_{x \geq 1} x \, P(\al (b_{k,t} E_{k,t})^{'} \geq x \mid \F_{t-1}, A_t) \right].
\end{align*}
The second inequality follows from the fact that $E_{k,t}^{b} \leq b_{k,t} E_{k,t}$. The fourth line follows from $\frac{1}{\al} P(\al (b_{k,t} E_{k,t})^{'} \geq 1)$ is always an upper bound for $\frac{y}{\al} P(\al (b_{k,t} E_{k,t})^{'} \geq y)$ for any $0 \leq y \leq 1$. Using the fact $\al (b_{k,t} E_{k,t})^{'} \in \frac{K}{[K]}$, we obtain the equivalent representation as the following:
\begin{align*}
\text{FDR}_t  & = \frac{1}{K} \sum_{k \in N_{0,t}} \E_{H_{0,k,t}} \Bigg[ \max_{y \in \frac{K}{[K]}} y P_{H_{0,k,t}}(\al \, (b_{k,t} E_{k,t})^{'} \geq y \mid \F_{t-1}, A_t) \Bigg] 
\\ &  \leq \frac{1}{K} \sum_{k \in N_{0,t}} \E_{H_{0,k,t}} \Bigg[ \underset{1 \leq y \leq K}{\max} \, y P_{H_{0,k,t}} \Big( b_{k,t} E_{k,t} \geq \frac{y}{\al} \mid \F_{t-1}, A_{t} \Big) \Bigg]
\\ &  \leq \frac{1}{K} \sum_{k \in N_{0,t}} \al \E_{H_{0,k,t}} \left[ \E_{H_{0,k,t}} \left[ E_{k,t} \mid \F_{t-1}, A_t \right] \right]
\\ &  = \frac{\al}{K} \sum_{k \in N_{0,t}} \E_{H_{0,k,t}} [E_{k,t}]
\\ &  \leq \al.
\end{align*}
The first inequality follows from $(b_{k,t} E_{k,t})^{'} \leq b_{k,t} E_{k,t}$ and the second follows from the condition given in \eqref{e:tipdset}, with $b_{k,t} \in B_2$ from Theorem \ref{th:conservative}, which is proved in Theorem \ref{th:conservative}. This completes the proof of the result for SHEREF-TIPD. The bound on \text{maxFDR} follows by taking supremum of $\text{FDR}_t$ over all values of $t \geq 1$. %
\end{proof}

\begin{proof}[Proof of Theorem \ref{th:conservative}]
We first establish that $B_1$, $B_2$ are non empty and $b_{k,t}^{GD} \in B_1$, $b_{k,t}^{TIPD} \in B_2$. Consider the inequality in \eqref{e:gd2} by multiplying $\al$ on both sides:
\begin{equation}
\E_{H_{0,k,t}} \Bigg[\al b  E_{k,t} 1_{\left(\al b E_{k,t} \geq 1 \right)} \Big| \F_{t-1}, A_t \Bigg] \\ \leq \al \E_{H_{0,k,t}}[E_{k,t} \mid \F_{t-1}, A_{t}]. 
\end{equation}
The inequality is satisfied when $b=1$, implying $B_1$ is nonempty. Moreover, $b \to \E_{H_{0,k,t}} \Bigg[\al b  E_{k,t} 1_{\left(\al b E_{k,t} \geq 1 \right)} \Big| \F_{t-1}, A_t \Bigg]$ is nondecreasing in $b$, so $b_{k,t}^{GD} \in B_1$ under \eqref{e:gd2}. Similarly, $B_2$ is nonempty since $1 \in B_2$, and $b \to \underset{1 \leq y \leq K}{\max} \, y P_{H_{0,k,t}} \left( b E_{k,t} \right)$ is nondecreasing in $b$. Thus, $b_{k,t}^{TIPD} \in B_2$.
The first part of the inequality, $1 \leq b_{k,t}^{GD}$ follows from $1 \in B_1$ and the fact that $\E_{H_{0,k,t}} \Bigg[\al b  E_{k,t} 1_{\left(\al b E_{k,t} \geq 1 \right)} \Big| \F_{t-1}, A_t \Bigg]$ is nondecreasing in $b$.
Next, we prove $b_{k,t}^{GD} \leq b_{k,t}^{TIPD}$. For any $y \geq 1$, we observe:
\begin{align*}
\E_{H_{0,k,t}} \Bigg[\al b  E_{k,t} 1_{\left(\al b E_{k,t} \geq 1 \right)} \Big| \F_{t-1}, A_t \Bigg]
& \geq \E_{H_{0,k,t}} \Bigg[\al b  E_{k,t} 1_{\left(\al b E_{k,t} \geq y \right)} \Big| \F_{t-1}, A_t \Bigg]
\\& \geq y \, P_{H_{0,k,t}} (\al b E_{k,t} \geq y \mid \F_{t-1}, A_t). 
\end{align*}
Taking maximum over $1 \leq y \leq K$ on both sides, we obtain:
\begin{equation} \label{e:gdtipd}
\E_{H_{0,k,t}} \Bigg[\al b  E_{k,t} 1_{\left(\al b E_{k,t} \geq 1 \right)} \Big| \F_{t-1}, A_t \Bigg]
\\ \geq \max_{1 \leq y \leq K} y \, P_{H_{0,k,t}} (\al b E_{k,t} \geq y \mid \F_{t-1}, A_t).
\end{equation}
This implies $B_1 \subseteq B_2$. Since $b_{k,t}^{GD} \in B_1$ and $b_{k,t}^{TIPD} \in B_2$, it follows that $b_{k,t}^{GD} \leq b_{k,t}^{TIPD}$ for $\de_1$ and $\de_2$ satisfying $\de_1 = \de_2 \leq \sup B_2 - \sup B_1$. 
\end{proof}

\begin{proof}[Proof of Theorem \ref{th:validex1}]
The Assumption \ref{con:restriction} is either part of the data generation mechanism or is directly satisfied for $A_t=[K]$. Thus for the first part of the the theorem, it is sufficient to analyze the conditional density function of $(x_{k,t} \mid \F_{t-1},A_t)$ and verify that it satisfies Assumption \ref{ass2}. %
We do this separately for each Example.
For each $t$, $k \in \Zb_{+}$, we have:
\begin{enumerate}
\item{Example \ref{ex:indep}: 
\begin{equation*}
(X_{k,t} \mid \F_{t-1}, A_t) \sim
\begin{cases}
f_{0,k,t}(x_{k,t}) & t \leq \tau_k
\\ f_{1,k,t}(x_{k,t}) & t > \tau_k,
\end{cases}
\end{equation*}
where we used the fact that $X_{k,t}$ is independent with $\F_{t-1}$ and $A_t$ under this data generation mechanism.
Also, the support condition follows from the assumption.
}
\item{Example \ref{ex:condep}: 
\begin{equation*}
(X_{k,t} \mid \F_{t-1}, A_t) \\ \sim
\begin{cases}
\int_{\mathbb{R}^{d_1}} f_{\ep_{k,t}}(x_{k,t} - h_{0,k,t}(z)) f_{Z_{t}}(z)\, dz  & t \leq \tau_k
\\ \int_{\mathbb{R}^{d_1}} f_{\ep_{k,t}}(x_{k,t} - h_{1,k,t}(z)) f_{Z_{t}}(z)\, dz & t > \tau_k.
\end{cases}
\end{equation*}
Also, the support for both densities is $\mathbb{R}^{d_2}$, which follows from the support of $\ep_{k,t}$ is defined as $\mathbb{R}^{d_2}$.}
\item{Example \ref{ex:depstrm}: 
\begin{equation*}
(X_{k,t} \mid \F_{t-1}, A_t) \\ \sim
\begin{cases}
f_{\ep_{k,t}}(x_{k,t} - h_{0,k,t}(y_{k,t-1}))  & t \leq \tau_k
\\ f_{\ep_{k,t}}(x_{k,t} - h_{1,k,t}(y_{k,t-1}))  & t > \tau_k.
\end{cases}
\end{equation*}
Also, the support for both densities is $\mathbb{R}^{d_2}$, which follows from the support of $\ep_{k,t}$ is given as $\mathbb{R}^{d_2}$.}
\item{Example \ref{ex:fixindex}: 
\begin{equation*}
(X_{k,t} \mid \F_{t-1}, A_t) \\ \sim
\begin{cases}
f_{0,k,t}(x_{k,t} \mid (x_{k,s})_{k \in [K], 1 \leq s \leq t-1})  & t \leq \tau_k
\\ f_{1,k,t}(x_{k,t} \mid (x_{k,s})_{k \in [K], 1 \leq s \leq t-1}) & t > \tau_k.
\end{cases}
\end{equation*}
Also, the support for both densities is $\mathbb{R}^{d}$, which follows from the support of $\ep_{k,t}$ is defined as $\mathbb{R}^{d}$.}
\end{enumerate}

For the second part of the theorem, we first establish time independence. The assumptions of Examples \ref{ex:indep} and \ref{ex:condep} ensure that, for every $k$, $X_{k,t}$ is independent of $\F_{t-1}$, thereby satisfying the time independence condition. Furthermore, when combined with Assumption \ref{con:restriction}, this implies for all $k,t \in \Zb_{+}$ and $l = 0,1$, $f_{l,k,t}(X_{k,t} \mid \F_{t-1}, A_t) = f_{l,k,t}(X_{k,t})$ almost surely. In Example \ref{ex:condep}, this guarantees $\tilde{L}_{k,t}$ defined in \eqref{e:lr} is well-defined for $k \in \Zb_+$, and the existence of a measurable function $\xi_{k,t}$ such that, $\tilde{L}_{k,t} = \xi_{k,t}(X_{k,t})$. Moreover, from the construction of \eqref{e:meanft}, $\tilde{L}_{k,t}$ is either a nondecreasing or nonincreasing function of $X_{k,t}$ for all $k \in \Zb_{+}$. In Example \ref{ex:indep}, the existence of such a monotone function $\xi_{k,t}$ is ensured by the assumption.

Now, we state a result from the \cite{sarkar2002}, to establish positive dependence.

Let $\tilde{X} = (X_1,\cdots,X_{K_0})$ for some finite $K_0 \in \mathbb{R}$. $\tilde{X}$ is jointly independent such that $\tilde{X}$ has a joint density $f$ with respect to some baseline measure or follows multivariate normal distribution $N(\delta, \Sig)$, with some $\delta \in \mathbb{R}^{K_0}$, and $\Sig_{i,j} \geq 0$ for $i \neq j$. Then $\tilde{X}$ is positive regression dependent on a subset (PRDS) on any subset of $\tilde{X}$. That is, for any increasing set $A \subset \mathbb{R}^{K_0}$, and for any $X^{*} \subset \tilde{X}$, the conditional probability $P(\tilde{X} \in A \mid X^{*} = x^{*})$ is elementwisely nondecreasing in $x^{*}$.

Combining the above facts, we conclude that for any increasing set $A \subset \mathbb{R}^{K_0}$, the conditional probability $P(\tilde{X} \in A \mid X_i = x)$ is nondecreasing in $x \in \mathbb{R}$ for any $X_i \in \tilde{X}$ when $\tilde{X}$ is jointly independent or follows a multivariate normal distribution with nonnegative correlations.

Assuming $d=d_2=1$ in Examples \ref{ex:indep} and \ref{ex:condep}, let $\tilde{X}_{t}$ be any subset of $\{X_{k,t}\}_{k \in \Zb_{+}}$ of size at most $K$ for each $t \in \Zb_{+}$. By the assumptions in Example \ref{ex:indep} and \ref{ex:condep}, $\tilde{X}_t$ consists of jointly independent variables in Example \ref{ex:indep}, and follows a positively correlated Gaussian distribution in Example \ref{ex:condep}. Applying the stated facts above, we conclude that since $\tilde{X}_{t}$ is either jointly independent or follows a positively correlated Gaussian distribution, $P(\tilde{X}_{t} \in A \mid X_{i,t} = x)$ is nondecreasing in $x$ for any $X_{i,t} \in \tilde{X}_t$, in both examples.
By replacing the increasing set $A$ with a decreasing set $B$, $P(\tilde{X}_{t} \in B \mid X_{i,t} = x)$ is nonincreasing in $x$.

Next, we show that $P(\tilde{X}_{t} \in B \mid X_{i,t} = x)$ being nonincreasing in $x$ implies that $P(\tilde{X}_{t} \in B \mid X_{i,t} \geq x)$ is nonincreasing in $x$. For each $i$, $t$ and any $x_1 > x_2$, 
define $\gam_t = P(X_{i,t} \geq x_1 \mid X_{i,t} \geq x_2)$. Then, we obtain the following bound: 
\begin{align*}
P(\tilde{X}_t \in B \mid X_{i,t} \geq x_2) 
&= \E[P(\tilde{X}_t \in B \mid X_{i,t}) \mid X_{i,t} \geq x_2]
\\ &= (1-\gam_t) \, \E[ P(\tilde{X}_{t} \in B \mid X_{i,t}) \mid x_2 \leq X_{i,t} < x_1]
+ \gam_t \, \E[ P(\tilde{X}_t \in B \mid X_{i,t}) \mid X_{i,t} \geq x_1]
\\ & \geq (1 - \gam_t) \, \E[P(\tilde{X}_t \in B \mid X_{i,t}) \mid X_{i,t} \geq x_1] 
+ \gam_t \, \E[P(\tilde{X}_{t} \in B \mid X_{i,t}) \mid X_{i,t} \geq x_1]
\\ & = P(\tilde{X}_t \in B \mid X_{i,t} \geq x_1).
\end{align*}
The first inequality follows from the assumption that $P(\tilde{X}_{t} \in B \mid X_{i,t} = x)$ is nonincreasing in $x$. The final equality follows from $\sig(\{X_{i,t} \geq x_1\}) \subset \sig(X_{i,t})$. Thus, if $\tilde{L}_{k,t} = \xi_{k,t}(X_{k,t})$ for nondecreasing function $\xi_{k,t}$, the same property holds. For a nonincreasing function $\xi_{k,t}$, we apply the same arguments to $-\tilde{X}_{t}$, proving that $P({L}_t^{*} \in B \mid \tilde{L}_{i,t} \geq x)$ is nonincreasing function in $x$, where ${L}_{t}^{*}$ is any finite subset of $\{\tilde{L}_{k,t}\}_{k \in \Zb_{+}}$ of size at most $K$ and $\tilde{L}_{i,t} \in L_{t}^{*}$. 

Finally, by the existence of a measurable function $\xi_{k,t}$ along with $X_{k,t}$ independent of $\sig(\F_{t-1}, A_t)$, %
we conclude that for all $k \in \Zb_{+}$, and $\tilde{L}_{k,t} \in L_{t}^{*}$,
\begin{equation}
P(L_t^{*} \in B \mid \tilde{L}_{k,t} \geq x) = P(L_t^{*} \in B \mid \tilde{L}_{k,t} \geq x, \F_{t-1}, A_t).
\end{equation}
Thus, $P(L_t^{*} \in B \mid \tilde{L}_{k,t} \geq x, \F_{t-1}, A_t)$ is nonincreasing in $x$, which proves the theorem.
\end{proof}

\section{Lemmas to Prove Theorem \ref{th:gdprocedure}} \label{appendB}
\noindent In this section, we provide supporting lemmas to prove Theorem \ref{th:gdprocedure}. These lemmas are slight modifications of results in \cite{blanchard2008} and \cite{wang2022} to account for a dynamically changing sensor network in our problem.
\begin{lemma} \label{pf:lemma0}
For each $t \geq 1$, define: 
\begin{equation*}
Q_t(u) = |\{ k \in \Zb_{+} \mid E_{k,t}^{b} \geq u \}| \vee 1,
\end{equation*}
and the detection threshold:
\begin{equation*}
u_{\al, t} = \inf \Big\{ u \in [0, \infty) \mid u Q_t(u) \geq \frac{K}{\al} \Big\}.
\end{equation*}
Then, Algorithm \ref{alg:alg1} applied to $\{E_{k,t}^{b}\}_{k\in A_t}$ under the target level $\al$ detects a change in $k$-th sensor if and only if 
\begin{equation*}
E_{k,t}^{b} \geq u_{\al, t},
\end{equation*}
where $u_{\al,t}Q_t(u_{\al,t}) = \frac{K}{\al}$.
\begin{proof}
Without loss of generality, we assume $A_t = \{1, \dots, |A_t|\}$. This lemma slightly extends Proposition 1 in \cite{wang2022}. Specifically, under Assumption \ref{con:restriction}, we have $|\{E^b_{k,t}: k \in A_t\}| = |A_t| \leq K$ in our setting, whereas in \cite{wang2022}, the number of e-values is exactly $K$. However, the exact number of available e-values in \cite{wang2022} is solely utilized to ensure the 
\[
\frac{u Q_t(u)}{K} \leq u
\]
holds for all $u$ in its proof. In our case, because $Q_t(u) \leq |A_t| \leq K$, this inequality remains valid. Hence, the lemma is established by following the same arguments used in \cite{wang2022}.
\end{proof}
\end{lemma}

\begin{lemma} \label{pf:lemma1}
Let $(W,V)$ be a couple of nonnegative random variables satisfying the following.
\begin{enumerate}
\item{There exists some measurable function $h(\cdot)$ and possibly dependent random variables $Z$ and $Y$ such that $V=h(Z,Y)$.}
\item{$P(W \leq q \mid Z=z) \leq q$ for all $q \in [0,1]$.}
\item{$w \rightarrow P(V \leq v \mid W \leq w, Z=z)$ is nondecreasing function of $w$.}
\end{enumerate}
Then, almost surely,
\begin{equation}
\E \left[ \frac{1_{(W \leq V)}}{V} \Big| Z\right] \leq 1.
\end{equation}
\begin{proof}
This lemma slightly extends Lemma 3.2 from \cite{blanchard2008} by replacing unconditional probabilities with  conditional probabilities given a random variable $Z$. The proof relies on two conditions adapted to this conditional setting: (1) $P(W \leq q) \leq q$ for $q \in [0,1]$ is replaced with $P(W \leq q \mid Z=z) \leq q$ for $q \in [0,1]$ and (2) $w \to P(V \leq v \mid W \leq w)$ is nondecreasing in $w$ is replaced by $w \to P(V \leq v \mid W \leq w, Z=z)$ is nondecreasing in $w$. With these adaptations, the proof follows the same reasoning as in Lemma 3.2 of \cite{blanchard2008}.
\end{proof}
\end{lemma}

Building on the first lemma, we state and prove the second lemma. While the proof of the second lemma has some similarities with Lemma 1 in \cite{wang2022}, our setting necessitates a more careful treatment of conditional probabilities. Therefore, for completeness and clarity, we provide a self-contained proof.

\begin{lemma} \label{pf:lemma2}
Let $\phi_t$ be a non-negative function measurable to $\F_t$, and $I_{\phi_t}$ denote its range. Define $E_{k,t}^{'}= \sup\{y \in I_{\phi_t} \cup \{0\} \mid y \leq E_{k,t} \}$. Additionally, define $\tilde{L}_{k,t} = \frac{f_{1,k,t}(X_{k,t} \mid \F_{t-1} ,A_t)}{f_{0,k,t}(X_{k,t} \mid \F_{t-1}, A_t)}$, $\tilde{\Lb}_{t} = (\tilde{L}_{k,t})_{k \in A_t}$, and $E_{k,t}^{b} = 1_{(k \in A_t)}b_{k,t}E_{k,t}$, where $b_{k,t}$ is defined from \eqref{e:tipd2}.
Suppose the following conditions hold:
\begin{enumerate}
\item{Assumption \ref{ass2}, \ref{con:restriction}, and the TIPD in \ref{def:tipd} are satisfied.}
\item{For each $t \geq 1$, define the collection of random variables as $\Qb_t$, where $\Qb_{t} = ((S_{k,t})_{k \in \Zb_+}, A_{t+1})$.}
\item{For each $t \geq 1$, $\phi_t(\Qb_{t-1}, \tilde{\Lb}_t)$ is nonincreasing in each component of $\tilde{\Lb}_{t}$.}
\end{enumerate}
Then, for all $k \in \Zb_{+}$ and $t \geq 1$, the following inequality holds almost surely:
\begin{equation}
\E \left[ \phi_t(\Qb_{t-1}, \tilde{\Lb}_{t}) 1_{(E_{k,t}^{b} \geq \phi_t(\Qb_{t-1}, \tilde{\Lb}_{t}))} \Big| \F_{t-1}, A_t \right] \\  \leq \sup_{x \geq 0} x P( (E_{k,t}^{b})^{'} \geq x \mid \F_{t-1}, A_t).
\end{equation}
\begin{proof}
For given $A_t$, if $k \notin A_t$, the inequality holds, since $E_{k,t}^b = 0$. Consider any $k \in A_t$. Under the time independence in Assumption \ref{def:tipd} and Assumption \ref{con:restriction}, for all $i \in \{0,1\}$ there exist a measurable function $f_{i,k,t}$ such that the following holds almost surely:
\begin{equation} 
f_{i,k,t}(X_{k,t} \mid \F_{t-1}, A_t) 
= f_{i,k,t}(X_{k,t} \mid \F_{t-1}) = f_{i,k,t}(X_{k,t}). 
\end{equation}
Consequently, there exists a measurable function $\xi_{k,t}$ such that $\tilde{L}_{k,t} = \xi_{k,t}(X_{k,t})$, and $\tilde{L}_{k,t}$ is independent of $\sig(\F_{t-1}, A_t)$.

Recall that $E_{k,t}^{b} = 1_{(k \in A_t)} b_{k,t} E_{k,t}$, where $b_{k,t}$ is $\F_{t-1}$ measurable, and $E_{k,t}$ is $\sig((S_{k,t-1})_{k \in \Zb_+}, A_{t}, \tilde{L}_{k,t})$ measurable from \eqref{e:iterative}. This implies there exist a function $\psi$ measurable to $\sig(\Qb_{t-1}, \tilde{L}_{k,t})$, such that $(E_{k,t}^{b})^{'} = \psi(\Qb_{t-1}, \tilde{L}_{k,t})$. Since $E_{k,t}$ is a nondecreasing function of $\tilde{L}_{k,t}$, $(E_{k,t}^{b})^{'}$ is also a nondecreasing function of $\tilde{L}_{k,t}$. We now introduce an independent copy $\tilde{O}_{k,t}$ of $\tilde{L}_{k,t}$, such that it is also independent of $\sig(\F_{t-1}, A_t)$. 
Define the function: 
\begin{equation*}
g_{k,t}(x)=P((\tilde{E}_{k,t}^{b})^{'} \geq x \mid \F_{t-1}, A_t, \tilde{L}_{k,t}),
\end{equation*}
where $(\tilde{E}_{k,t}^{b})^{'} = \psi(\Qb_{t-1}, \tilde{O}_{k,t})$. Following a similar reasoning, $(\tilde{E}_{k,t}^{b})^{'}$ is nondecreasing function of $\tilde{O}_{k,t}$. Then, defining $P_{k,t} = g_{k,t}((E_{k,t}^{b})^{'})$, we almost surely obtain: 
\begin{equation*}
P(P_{k,t} \leq q \mid \F_{t-1}, A_t) \leq q.
\end{equation*}
This holds from Lemma \ref{pf:lemma5}, by letting $Z = \Qb_{t-1}$, $X = \tilde{O}_{k,t}$, $Y = \tilde{L}_{k,t}$, and $h(\cdot) = \psi(\cdot)$, noting that $\Qb_{t-1}$ is $\sig(\F_{t-1}, A_t)$ measurable. Notice that $P_{k,t}$ is nonincreasing function of $\tilde{L}_{k,t}$ followed by $(E_{k,t}^{b})^{'}$ is nondecreasing function of $\tilde{L}_{k,t}$ from \eqref{e:iterative}.

Next, by definition, $\phi_t$ is nonincreasing function in each component of $\tilde{\Lb}_t$. Since $(g_{k,t} \circ \phi_t)(\Qb_{t-1}, \tilde{\Lb}_t)$ is also nondecreasing in $\tilde{\Lb}_t$, $[0,y]$ for any $y \geq 0$ is decreasing set, and given the positive dependence in TIPD, it follows that for $k \in A_t$, the mapping $x \rightarrow P( (g_{k,t} \circ \phi_t)(\Qb_{t-1}, \tilde{\Lb}_t) \leq y \mid \tilde{L}_{k,t} \geq x, \F_{t-1}, A_t)$ is nonincreasing on $[0,\infty)$. Similarly, since $P_{k,t}$ is a nonincreasing function of $\tilde{L}_{k,t}$ we conclude that for any $k \in A_t$, $z \rightarrow P( (g_{k,t} \circ \phi_t)(\Qb_{t-1}, \tilde{\Lb}_t) \leq y \mid P_{k,t} \leq z, \F_{t-1}, A_t)$ is nondecreasing on $z \in [0,1]$. Now, for all $k \in A_t$:
\begin{align*}
\E \left[ \phi_t(\Qb_{t-1}, \tilde{\Lb}_{t}) 1_{(E_{k,t}^{b} \geq \phi_t(\Qb_{t-1}, \tilde{\Lb}_{t}))} \Big| \F_{t-1}, A_t \right] & = \E \left[ \phi_t(\Qb_{t-1}, \tilde{\Lb}_{t}) 1_{((E_{k,t}^{b})^{'} \geq \phi_t(\Qb_{t-1}, \tilde{\Lb}_{t}))} \Big| \F_{t-1}, A_t \right]
\\ &\leq \E \left[ \phi_t(\Qb_{t-1}, \tilde{\Lb}_{t}) 1_{(P_{k,t} \leq (g_{k,t} \circ \phi_t)(\Qb_{t-1}, \tilde{\Lb}_{t}))} \mid \F_{t-1}, A_t \right]
\\ &= \E \Big[ \phi_t(\Qb_{t-1}, \tilde{\Lb}_{t}) 1_{(P_{k,t} \leq (g_{k,t} \circ \phi_t)(\Qb_{t-1}, \tilde{\Lb}_{t}))}
\\ & \hspace{1cm} \times \frac{P(({\tilde{E}}_{k,t}^{b})^{'} \geq \phi_t(\Qb_{t-1}, \tilde{\Lb}_{t}) \Big | \F_{t-1}, A_t, \tilde{L}_{k,t})}{ (g_{k,t} \circ \phi_t) (\Qb_{t-1}, \tilde{\Lb}_{t})} \Big| \F_{t-1}, A_t \Big]
\\ & \leq \E[ \sup_{x \geq 0} x P( (\tilde{E}_{k,t}^{b})^{'} \geq x \mid \F_{t-1}, A_t, \tilde{L}_{k,t}) \mid \F_{t-1}, A_t]
\\ & \hspace{1cm} \times \E \left[ \frac{1_{(P_{k,t} \leq (g_{k,t} \circ \phi_t)(\Qb_{t-1}, \tilde{\Lb}_{t}))}}{ (g_{k,t} \circ \phi_t) (\Qb_{t-1}, \tilde{\Lb}_{t})} \Big| \F_{t-1}, A_t \right]
\\ & \leq \E[ \sup_{x \geq 0} x P( (\tilde{E}_{k,t}^{b})^{'} \geq x \mid \F_{t-1}, A_t, \tilde{L}_{k,t}) \mid \F_{t-1}, A_t]
\\ & = \E[ \sup_{x \geq 0} x P( (\tilde{E}_{k,t}^{b})^{'} \geq x \mid \F_{t-1}, A_t) \mid \F_{t-1}, A_t]
\\ & = \sup_{x \geq 0} x P((E_{k,t}^{b})^{'} \geq x \mid \F_{t-1}, A_t),
\end{align*}
where the third inequality follows from Lemma \ref{pf:lemma1}, with $V = (g_{k,t} \circ \phi_t)(\Qb_{t-1}, \tilde{\Lb}_{t})$, and $W = P_{k,t}$. The third equality follows from $(\tilde{E}_{k,t}^{b})^{'} = \psi(\Qb_{t-1}, \tilde{O}_{k,t})$, and $\tilde{O}_{k,t}$ is independent of $\tilde{L}_{k,t}$. The final equality follows from $\tilde{E}_{k,t}^{b}$ and $E_{k,t}^{b}$ have identical distribution. This completes the proof.
\end{proof}
\end{lemma}

\begin{lemma} \label{pf:lemma5}
Let $X$ be a random variable and $Y$ an independent copy of $X$, such that $X, Y \in \mathbb{R}$, $X$, $Y$ are independent of a random variable $Z \in \mathbb{R}^d$ for some $d$.  Let $h:\mathbb{R}^d\times \mathbb{R} \to \mathbb{R}$ be a measurable function such that for fixed $Z$, $x \to h(z,x)$ and $y \to h(z,y)$ are nondecreasing in $x$ and $y$ respectivley. Then, the following holds for any $q \in [0,1]$:
\begin{equation}
P( P( h(Z,X) \geq h(Z,Y) \mid Z,Y ) \leq q \mid Z) \leq q.
\end{equation}
\begin{proof}
From the nondecreasing property of measurable function $h$ in its second argument, and using the independence of $X$ and $Y$ from $Z$, we almost surely obtain:
\begin{align*}
P(h(Z,X) \geq h(Z,Y) \mid Z,Y ) &\geq P(X \geq Y \mid Z,Y)
\\& = P(X \geq Y \mid Y).
\end{align*}
Applying this inequality and using the independence of $X$ and $Y$ from $Z$, we derive the following upper bound:
\begin{align*}
P( P( h(Z,X) \geq h(Z,Y) \mid Z,Y) \leq q \mid Z)
& \leq P( P( X \geq Y \mid Y) \leq q \mid Z)
\\& = P( P( X \geq Y \mid Y) \leq q).
\end{align*}
Now define $F_{X}(x) = P( X \leq x )$ and its left limit $F_{X}(x^{-}) = \lim_{\epsilon \to 0^+} F_{X}(x - \epsilon)$. Rewriting the probability,
\begin{equation*}
P( P ( X \geq Y \mid Y ) \leq q ) = P (F_{X}(Y^{-}) \geq 1-q).
\end{equation*}
Consider three cases:
\begin{enumerate}
\item{Suppose there exists $y_0$ such that $F_{X}(y_0) = 1-q$, and define $y_{1-q}$ such that $y_{1-q} = \inf\{ y : F_{X}(y) = 1-q \}$. If $1-q = F_{X}(y_{1-q}) = F_{X}(y_{1-q}^{-})$, then:
\begin{align*}
P (F_{X}(Y^{-}) \geq 1-q) &=
P (Y^- \geq y_{1-q}) 
\\& \leq P ( Y \geq y_{1-q})
\\& = 1 - F_{X}(y_{1-q}) 
\\& = q.
\end{align*}}

\item{Suppose there exists $y_0$ such that $F_{X}(y_0) = 1-q$, and $1-q = F_{X}(y_{1-q}) > F_{X}(y_{1-q}^{-})$, then:
\begin{align*}
P (F_{X}(Y^{-}) \geq 1-q) & \leq P(Y > y_{1-q}) \\& = 1 - F_{X}(y_{1-q}) \\& = q.
\end{align*}}

\item{If no such $y_{0}$ exists, choose $y_0$ such that $F_{X}(y_{0}^-) < 1-q < F_{X}(y_{0})$, then:
\begin{align*}
P(F_{X}(Y^{-}) \geq 1-q) &\leq P(Y > y_0) \\& = 1 - F_{X}(y_0) \\& \leq q.
\end{align*}}
\end{enumerate}
Thus, the inequality holds in all cases, completing the proof.
\end{proof}
\end{lemma}

\bibliographystyle{IEEEtran}
\bibliography{IEEEabrv,reference}
\end{document}